\documentclass[nojss]{jss}

\usepackage{orcidlink,thumbpdf,lmodern}
\usepackage[normalem]{ulem} 
\usepackage{framed}
\usepackage{amsmath}
\usepackage{bm}

\author{Maria De Martino\\University of Udine\\Karolinska Institutet 
   \And Federico Triolo\\Karolinska Institutet
   \And Adrien Perigord\\Karolinska Institutet
   \AND Alice Margherita Ornago\\Karolinska Institutet
   \And Davide Liborio Vetrano\\Karolinska Institutet
   \AND Caterina Gregorio\\Karolinska Institutet}
\Plainauthor{M. De Martino, F. Triolo, A. Perigord, A. M. Ornago, D. L. Vetrano, C. Gregorio}

\title{MixMashNet: An \proglang{R} Package for Single and Multilayer Networks}
\Plaintitle{MixMashNet: An R Package for Single and Multilayer Networks}
\Shorttitle{MixMashNet: An \proglang{R} Package for Networks}

\Abstract{
 The \proglang{R} package \pkg{MixMashNet} provides an integrated framework for estimating and analyzing single and multilayer networks using Mixed Graphical Models (MGMs), accommodating continuous, count, and categorical variables. In the multilayer setting, layers may comprise different types and numbers of variables, and users can explicitly impose a predefined multilayer topology. Bootstrap procedures are implemented to quantify sampling uncertainty for edge weights and node-level centrality indices. In addition, the package includes tools to assess the stability of node community membership and to compute community scores that summarize the latent dimensions identified through network clustering. \pkg{MixMashNet} also offers interactive \proglang{Shiny} applications to support exploration, visualization, and interpretation of the estimated networks.
}

\Keywords{Network, Multilayer Network, Mixed Graphical Models, Bootstrap Stability, \proglang{R}}
\Plainkeywords{JSS, style guide, comma-separated, not capitalized, R}

\Address{
  Maria De Martino\\
  Division of Medical Statistics\\
  University of Udine\\
  Udine, Italy\\
  \emph{and}\\
  Aging Research Center\\
  Department of Neurobiology, Care Sciences and Society\\
  Karolinska Institutet and Stockholm University\\
  Stockholm, Sweden\\
  E-mail: \email{maria.demartino@uniud.it}
\vspace{1em}\\
  Federico Triolo\\
  Aging Research Center\\
  Department of Neurobiology, Care Sciences and Society\\
  Karolinska Institutet and Stockholm University\\
  Stockholm, Sweden\\
  E-mail: \email{federico.triolo@ki.se}
  \vspace{1em}\\
  Adrien Perigord\\
  Aging Research Center\\
  Department of Neurobiology, Care Sciences and Society\\
  Karolinska Institutet and Stockholm University\\
  Stockholm, Sweden\\
  E-mail: \email{adrien.perigord@ki.se}
  \vspace{1em}\\
  Alice Margherita Ornago\\
  Aging Research Center\\
  Department of Neurobiology, Care Sciences and Society\\
  Karolinska Institutet and Stockholm University\\
  Stockholm, Sweden\\
  E-mail: \email{alice.margherita.ornago@ki.se}
  \vspace{1em}\\
  Davide Liborio Vetrano\\
  Aging Research Center\\
  Department of Neurobiology, Care Sciences and Society\\
  Karolinska Institutet and Stockholm University\\
  Stockholm, Sweden\\
  E-mail: \email{davide.vetrano@ki.se}
  \vspace{1em}\\
  Caterina Gregorio\\
  Aging Research Center\\
  Department of Neurobiology, Care Sciences and Society\\
  Karolinska Institutet and Stockholm University\\
  Stockholm, Sweden\\
  E-mail: \email{caterina.gregorio@ki.se}
}

\begin{document}

\section[Introduction]{Introduction} \label{sec:intro}

Networks provide a powerful framework for representing and analyzing complex systems across a wide range of scientific disciplines. By modeling observed variables as nodes and their relationships as edges, these models offer an intuitive way to capture multidimensional dependencies in high-dimensional data. Despite the rapid growth of the field, some methodological barriers have hindered the effective implementation of network-based statistical analyses on empirical data, with many approaches developed within specific application domains. In particular, many existing methods focus on single layer networks, where all variables are represented within a single level of analysis, such as networks constructed exclusively from molecular biomarkers or from patient-level clinical features. Even when multilayer structures are considered, the same set of variables is often replicated across layers, with each layer capturing a different type of relationship or dependency among them. Moreover, many existing methods rely on the assumption that all variables are of the same type, typically requiring nodes to be either all continuous or all categorical. However, real-world data often involve multiple layers of organization, each having different sets of nodes, possibly defined by a mix of categorical and continuous variables, requiring models that can jointly characterize both intra and interlayer dependencies. 
Furthermore, as network models grow in complexity, introducing sparsity and assessing the robustness of the estimated features becomes crucial. Procedures to evaluate the stability of these structures are a key component of network analysis, yet they are still not systematically integrated into methodological frameworks.

Several methods have been proposed in recent years to address some of these challenges.
The \proglang{R} package \pkg{EGAnet} \citep{golinoEGAnetExploratoryGraph2019} offers a framework for estimating psychological and psychometric single layer
networks, and implements bootstrap procedures to evaluate the stability of nodes within their communities. The \proglang{R} package \pkg{bootnet} \citep{epskampEstimatingPsychologicalNetworks2018} enables the assessment of stability of estimated single layer networks and related statistics, including edge weights and node centrality indices. The \proglang{R} package \pkg{mgm} \citep{haslbeckMgmEstimatingTimeVarying2020} offers a resampling function to refit the estimated network model on bootstrap samples; however, its current implementation focuses solely on single layer networks and does not encompass multilayer configurations.
The \proglang{R} packages \pkg{MuxViz} \citep{dedomenicoMuxVizToolMultilayer2015}, \pkg{mully} \citep{hammoudMullyPackageCreate2018} and \pkg{emln} \citep{frydmanPracticalGuidelinesEMLN2023a} allow for the visualization and analysis of multilayer networks, but they rely on pre-computed adjacency matrices and do not provide statistical estimation of network structure from the data. As a result, they also lack built-in methods for uncertainty quantification. 
In the \proglang{Python} ecosystem, several libraries have been developed for network analysis, \pkg{NetworkX} \citep{paper:hagberg:2008} for the single layer case, and \pkg{pymnet} \citep{nurmiPymnetPythonLibrary2024}, \pkg{multinetX} \citep{amatoOpinionCompetitionDynamics2017}, and \pkg{py3plex} \citep{skrljPy3plexToolkitVisualization2019} for the multilayer case. However, these approaches do not provide integrated procedures for statistical estimation of network models from data, nor for the assessment of uncertainty in estimated network quantities.

Given that existing software implementations focus on single layer networks, provide only visualization and manipulation tools, or offer procedures for assessing the stability of network metrics in isolation, the \proglang{R} package \pkg{MixMashNet} introduces an integrated framework for estimating both single and multilayer networks. It accommodates continuous, count, and categorical variables using Mixed Graphical Models (MGMs) through the \proglang{R} package \pkg{mgm} \citep{haslbeckMgmEstimatingTimeVarying2020}. Its regularization-based estimation procedures make it suitable for high-dimensional data settings, enabling the simultaneous estimation of multiple layers of interconnected variables while preserving sparsity and interpretability across the whole network. \pkg{MixMashNet} implements a unified bootstrap infrastructure to assess the stability and reliability of key network features, including community structures, edge weights, and node-level indices. Beyond classical centrality measures, it includes indices to determine nodes that act as connectors between communities or layers, while also handling singletons i.e. nodes not assigned to any community. Visualization features leveraging \proglang{ggplot2}, together with an interactive \proglang{Shiny} application, enable intuitive exploration and interpretation of the single and multilayer network models obtained.

The paper is organized as follows. In Section \ref{sec:background}, we provide an overview of the theoretical framework and methodological foundations underlying the construction of the \pkg{MixMashNet} package. Section \ref{sec:package} presents an overview of the package architecture.
Section \ref{sec:usage} illustrates the usage of the package on real-world datasets, first focusing on the single layer case and then extending to the multilayer framework. Finally, Section \ref{sec:summary} concludes by summarizing the main contributions and potential applications of the package.

\section{Background} \label{sec:background}

This section reviews the theoretical background underlying \pkg{MixMashNet}. We first introduce single and multilayer networks (Section~\ref{sec:singlemulti}) and MGMs (Section~\ref{sec:MGM}). We then describe intra and interlayer analyses (Sections~\ref{sec:intra} and \ref{sec:inter}), and conclude with the bootstrap framework used to assess the stability of the estimated quantities (Section~\ref{sec:boot}).

\subsection{Definition} \label{sec:singlemulti}

\subsubsection*{\textit{Single Layer Networks}}

Networks are typically represented as a collection of $n$ nodes (or vertices), denoted by $V$, and a collection of $l$ edges, denoted by $E$, which describe the connections between the nodes. Hence, a network can be expressed as $G = (V,E)$, where $E \subseteq V \times V$. A network can be encoded by an $n \times n$ adjacency matrix $Y$, where the presence of an edge between nodes $i$ and $j$, with $i \neq j$, is indicated by

\[
Y_{ij} =
\begin{cases}
w_{ij} & \text{if there is an edge connecting node } i \text{ to node } j,\\
0 & \text{otherwise}.
\end{cases}
\]

This definition applies to weighted  networks--such as those considered throughout this paper--where the entries $w_{ij}$ represent the strength of the edge between nodes $i$ and $j$.

A network is said to be undirected if $Y_{ij} = Y_{ji}$, meaning that the adjacency matrix is symmetric; otherwise, it is directed. We focus on the undirected case. 

\subsubsection*{\textit{Multilayer Networks}}

Many real-world systems require more complex representations, with dependencies across multiple levels. Multilayer networks can provide a more flexible framework, by enabling the modeling of both intra and interlayer relationships within a unified representation.
However, the concept of a multilayer network is not uniquely defined, with different formalizations introduced depending on the application. Throughout this paper, we adopt the definition proposed by \cite{kivelaMultilayerNetworks2014a}. 

We begin with the same set of nodes $V$. In addition, we define a set of $d$ layers 
\[
\alpha_1, \alpha_2, \ldots, \alpha_d.
\]

Nodes do not need to be present in each layer; therefore, we define 
\[
V_M \subseteq V \times \alpha_1 \times \alpha_2 \times \cdots \times \alpha_d
\]
as the set of permissible node-layer tuples. A tuple $ (u, \alpha_{1}, \alpha_{2} \ldots, \alpha_{d})$ denotes the presence of node $u$ in that set of layers $\alpha_{1}, \alpha_{2}. \ldots, \alpha_{d}$.

The edges are collected in the set $E_{M} \subseteq V_{M} \times V_{M}$. An edge connecting two nodes belonging to the same layer is called an \textit{intralayer} edge, whereas edges connecting nodes from different layers are termed \textit{interlayer} edges.

A multilayer network can thus be denoted by the quadruplet \[
M = (V_M, E_M, V, \pmb{\alpha}),
\]
where $\pmb{\alpha} = (\alpha_1, \ldots, \alpha_d)$.  
When $d = 1$ the framework reduces to the single layer case. 

The multilayer structure can be encoded in a \textit{multilayer adjacency tensor} $M_{u \alpha_{i}}^{v \alpha_{j}}$, with $u, v \in V$ and $i,j \in {1,\ldots,d}$, where each entry represents the edge weight between node $u$ in layer $\alpha_{i}$ and node $v$ in layer $\alpha_{j}$ \citep{dedomenicoMoreDifferentRealworld2023}.

In this paper, we focus on a specific class of multilayer networks:
\begin{itemize}
  \item layer-disjoint networks, in which each layer contains a different set of nodes; and  
  \item heterogeneous networks, in which nodes include variables of different data types (continuous, count, categorical).
\end{itemize}

\subsection{Mixed Graphical Models} \label{sec:MGM}

Mixed Graphical Models (MGMs) provide a flexible framework for estimating networks from data containing variables of any nature (continuous, count, binary, and multilevel categorical). For a comprehensive understanding of MGMs we refer the reader to \cite{haslbeckMgmEstimatingTimeVarying2020} and \cite{leeLearningStructureMixed2015}; here, we provide an overview tailored to our setting.

In MGMs, the dependency structure is represented by a graph $G = (V,E)$, where each node in $V$ corresponds to a variable and edges in $E$ reflect conditional associations. Let $X_{i}$ denote the random variable associated with node $i \in V$ and $\pmb{X} = (X_{1}, \dots, X_{p})$ the full set of variables. The joint distribution of $\pmb{X}$ belongs to the exponential family and is written in terms of interaction functions over cliques $C$ of variables up to a maximum order k. A clique $C$ is a subset of $V$ such that $(s,t) \in E$, for all $s, t \in C$, with $s  \not= t$. 
Let $\mathcal{C}$ denote the set  of all cliques. The joint log-density can be expressed as:
\[
\log p(x) = \sum_{C \in \mathcal{C}} \theta_{C}\phi_{C}(\pmb{X_{C}})-\Phi(\theta)
\]
where $\phi_{C}$ are sufficient statistics associated with the clique $C$, $\theta_{C}$ are the corresponding interaction parameters, $\pmb{X_{C}} = \{X_{s}: s \in C\}$, and $\Phi(\theta)$  is the log-normalizing constant.
For example, in the univariate Gaussian distribution with known variance $\sigma^{2}$, the sufficient statistic can be written as $\phi(X)= \frac{X}{\sigma}$.
The parameter $k$ specifies the maximum clique size (i.e., the highest interaction order allowed): when $k=2$, the graph contains only pairwise interactions, and absence of an edge implies conditional independence given all remaining variables.

Estimation is performed through a nodewise approach: for each variable $X_{s}$ a generalized linear model (GLM; \cite{nelderGeneralizedLinearModels1972}) is fitted using all the remaining variables (and interaction terms if $k>1$) as predictors, selecting an appropriate link based on the data type of $X_{s}$. To ensure sparsity and applicability in  high-dimensional settings, nodewise models are estimated by maximizing a penalized likelihood, with the regularization parameter selected through cross-validation (CV) or the extended Bayesian information criterion (EBIC), allowing control over the number of CV folds or the EBIC tuning parameter $\gamma$.
Setting $\gamma =0$ yields the classical BIC, while the default choice 
$\gamma = 0.25$ has been shown to substantially reduce false positives without markedly increasing false negatives \citep{foygelExtendedBayesianInformation2010}.
More generally, penalization can be formulated as an elastic net, allowing for any convex combination of $l_{1}$ and $l_{2}$-penalties through the mixing parameter $\alpha \in [0,1]$, with $\alpha = 1$ corresponding to the lasso (default). When multiple values of $\alpha$ are considered, the optimal value can be selected using either CV or EBIC.

Each potential edge appears in two nodewise models (predicting $X_{s}$ from $X_{r}$ and vice versa). 
The global edge weight is obtained by aggregating the retained coefficients from the two regressions, by averaging their absolute values. For categorical variables with more than two levels, multiple dummy-coded parameters contribute to the same pairwise interaction; in this case, their absolute values are averaged to obtain a single edge weight.
The AND-rule retains an edge only if it is selected in both regressions, whereas the OR-rule includes it if selected in at least one.

A sign is also assigned to each edge. For continuous variables, it reflects the direction of the conditional association after adjusting for all other nodes. When binary variables are involved, the sign can likewise be interpreted based on whether the presence of the binary condition increases or decreases the expected value of the other variable. For variables with more than two categories, no unique sign can be assigned because the interaction is represented by multiple category-specific parameters; therefore, the sign is set to positive by convention.

In this work, we rely on the \pkg{mgm} package, which provides an implementation of MGMs in the \proglang{R} environment. However, \pkg{mgm} does not allow users to restrict the set of predictors that a given node can use in its nodewise regression. This poses a limitation in the multilayer context, where—depending on the chosen network construction—not all layers (and consequently not all nodes) are allowed to be connected.
To address this issue, we introduce a masked MGM estimator. For each variable $X_{s}$, we define a set of allowed predictors $A_{s} \subset V$, and exclude all coefficients corresponding to disallowed interactions by imposing
\[
\theta_{s,r} = 0 \quad \text{if } r \notin A_s .
\]

This is implemented by removing the corresponding predictors from the nodewise design matrix before estimation. As a result, edge weights that are incompatible with the predefined multilayer architecture are prevented from entering the model, ensuring that the estimated network respects the intended topology without requiring post-hoc pruning.

\subsection{Intralayer Analyses} \label{sec:intra}

This section provides an overview of the selected analyses that can be performed after estimating a single layer network, with a focus on the methods implemented in the \pkg{MixMashNet} package. We describe general centrality indices for assessing node importance, community detection algorithms, community-related centrality measures, new centrality indices defined for nodes not assigned in communities, and community scores  derived from the identified communities. 

\subsubsection*{General Centrality Indices}

In network analysis, centrality indices quantify the relative importance of nodes within a network.
For weighted undirected graphs, as considered in this work, we rely on the definitions proposed by \cite{opsahlNodeCentralityWeighted2010} and \cite{robinaughIdentifyingHighlyInfluential2016}. Specifically:
\begin{itemize}
\item \textit{Strength}: sum of the absolute weights of all edges incident to the node.
\item \textit{Closeness}: inverse of the total shortest-path distance from a node to all other nodes. For this index the edges are considered with their absolute value.
\item \textit{Betweenness}: number of shortest paths between all pairs of nodes that pass through the target node. For this index the edges are considered with their absolute value.
\item \textit{Expected Influence}: sum of all incident edge weights to a node, preserving their signs.
\end{itemize}

\subsubsection*{Community Detection Algorithms}

Nodes in a network can often be grouped into clusters such that nodes within the same cluster are more densely connected to each other than to nodes outside the cluster. For a formal and comprehensive definition, we refer to the review by \cite{fortunatoCommunityDetectionGraphs2010}. These clusters are commonly referred to as \textit{communities}. 
In this section, we provide a brief overview of the main community detection algorithms implemented in the \pkg{MixMashNet} package. 

\begin{itemize}
\item \textit{Louvain algorithm} \citep{blondelFastUnfoldingCommunities2008a}: a modularity-based method that iteratively aggregates nodes into communities to maximise modularity.
\item \textit{Fast-Greedy algorithm} \citep{clausetFindingCommunityStructure2004}: a hierarchical agglomerative approach that builds a dendrogram by merging communities to increase modularity.
\item \textit{Walktrap algorithm} \citep{ponsComputingCommunitiesLarge2006}: based on the idea that short random walks tend to remain within the same community.
\item \textit{Infomap algorithm} \citep{rosvallMapsRandomWalks2008}: detects communities by optimising the description length of random-walk flows on the network.
\item \textit{Edge betweenness algorithm} \citep{newmanFindingEvaluatingCommunity2004}: identifies communities by iteratively removing edges with the highest betweenness centrality, causing the network to split into components.
\end{itemize}

\subsubsection*{Bridge Centrality Indices}

We present in this section community-related centrality indices. These indices are defined to quantify the extent to which individual nodes act as connectors between distinct communities. These measures originate in the psychometric network literature and were defined by \cite{jonesBridgeCentralityNetwork2021a}. Specifically:

\begin{itemize}
\item \textit{Bridge Strength}: sum of the absolute weights connecting a node to nodes belonging to other communities.
\item \textit{Bridge Closeness}: closeness centrality of a node computed only with respect to nodes in distinct communities. Distances are computed from absolute edge weights.
\item \textit{Bridge Betweenness}: number of shortest paths that start and end in different communities and pass through the node. Shortest paths are computed using absolute edge weights.
\item \textit{Bridge Expected Influence (1-step)}: the signed sum of all edges linking a node to nodes in other communities.
\item \textit{Bridge Expected Influence (2-step)}: the signed sum of all edges linking a node to nodes in other communities, including the indirect (two-step) effects propagated through one intermediate node.
\end{itemize}

\subsubsection*{Bridge Centrality Indices for excluded nodes}

In this section, we introduce the concept of \textit{excluded nodes}. 
Based on the stability analysis discussed in Section \ref{sec:boot}, some nodes may not be consistently assigned to any single community. This occurs when their community membership is not sufficiently stable across bootstrap replications. Since in this work we do not consider overlapping communities, such nodes are labelled as \textit{excluded}. 
Nevertheless, we remain interested in the potential role these nodes may play in connecting different communities, acting as bridges between distinct domains. To capture this role, we define \textit{bridge centrality indices for excluded nodes}, which mirror the bridge centrality indices introduced in the previous section. Specifically:

\begin{itemize}
\item \textit{Bridge Strength for excluded nodes}: sum of the absolute weights connecting an excluded node to nodes belonging to assigned communities.
\item \textit{Bridge Closeness for excluded nodes}: closeness centrality of an excluded node computed with respect to nodes in assigned communities. Distances are computed from absolute edge weights.
\item \textit{Bridge Betweenness for excluded nodes}: number of shortest paths that start and end in two different communities and pass through the excluded node. Shortest paths are computed using absolute edge weights.
\item \textit{Bridge Expected Influence (1-step) for excluded nodes}: signed sum of all edges linking an excluded node to nodes in assigned communities.
\item \textit{Bridge Expected Influence (2-step) for excluded nodes}: the signed sum of all edges linking an excluded node to nodes in assigned communities, including the indirect (two-step) effects.
\end{itemize}

\subsubsection*{Community Scores}


Recent work \citep{golinoEGAnetExploratoryGraph2019} has shown that community detection does more than simply partitioning a network into meaningful subgraphs: it can also reveal latent dimensions underlying the data. This observation motivates the construction of network scores—which in our paper we refer to as community scores—that provide, for each individual, a numerical summary of the extent to which the variables within a community are jointly expressed \citep{christensenEquivalencyFactorNetwork2021}.

To compute these scores, one first quantifies how strongly each node is connected to the other nodes in its community through a set of network loadings, defined as node strength restricted to intracommunity edges. Community scores are then obtained, for each individual $i$ and community $c$, as linear combinations of the observed variables:
\[
s_{ic} = \sum_{j=1}^{p} x_{ij} \, o_{jc},
\]
where $x_{ij}$ denotes the standardized observed value of variable $j$ for individual $i$, $p$ is the total number of variables, and $o_{jc}$ represents the loading of node $j$ in community $c$, with $o_{jc} = 0$ for variables not belonging to community $c$. The resulting scores provide one value per community for each individual in the dataset and can be interpreted as data-driven components summarizing the level of "activation" of each community within an individual. These scores make it possible to leverage the network structure to evaluate the added value of modeling the joint behavior of variables within a community—beyond their individual effects—when addressing etiological questions or conducting predictive analyses (e.g. involving clinical or biological endpoints). 

In the \pkg{MixMashNet} package, we implement the network-score procedure from \pkg{EGAnet},  adopting its revised formulation proposed by \cite{christensenRevisedNetworkLoadings2025a}, in order to compute community scores.

\subsection{Interlayer Analyses} \label{sec:inter}

In \pkg{MixMashNet}, the interlayer structure is not fixed a priori: users may explicitly specify which pairs of layers are allowed to be connected. These rules determine the set of admissible cross-layer edges that the MGM may estimate, and all interlayer metrics are computed with respect to this user–defined multilayer topology.

Furthermore, the \pkg{MixMashNet} package focuses on
interlayer relations and on their role in mediating cross-layer communication. 
To this end, we define layer-specific analogues of the general node-level
indices, computed with respect to interlayer edges only. In particular:

\begin{itemize}
\item \textit{Interlayer Strength}: sum of the absolute weights of all edges
  connecting a node to nodes belonging to different layers.
\item \textit{Interlayer Closeness}: closeness centrality of a node computed
  on the graph that includes only interlayer edges. Distances are computed from absolute edge weights
\item \textit{Interlayer Betweenness}: number of shortest paths, computed on
  the interlayer-only graph (with distances based on absolute edge weights),
  that pass through the node.
\item \textit{Interlayer Expected Influence}: signed sum of all interlayer
  edges incident to a node.
\end{itemize}

\subsection{Uncertainty Quantification} \label{sec:boot}

One of the key features of \pkg{MixMashNet} is the implementation of a comprehensive bootstrap framework to evaluate the uncertainty associated with all estimated network quantities, as well as the stability of detected communities. This framework adapts and extends ideas from two existing packages: \pkg{EGAnet}, which provides bootstrap procedures for community stability but is not designed for MGMs, and \pkg{bootnet}, which focuses on stability of network parameters but does not address community stability. Importantly, neither package supports multilayer networks. 

 Uncertainty in \pkg{MixMashNet} is quantified using empirical quantile regions computed from the non-parametric bootstrap replicates of each statistic. Because network estimation relies on regularized MGMs, the resulting intervals are 
not intended for null-hypothesis significance testing. Moreover, as discussed by \cite{epskampEstimatingPsychologicalNetworks2018}, bootstrapping centrality indices may lead to biased sampling distributions. Therefore, the reported $95\%$ quantile regions should be interpreted as descriptive summaries of sampling uncertainty rather than as formal confidence intervals around a true population parameter. 

With respect to community stability, the underlying idea is to quantify how consistently a node is allocated to the same community across the bootstrap replicates, by computing the proportion of bootstrap replications in which the node is assigned to the same community as in the original estimate. This approach was introduced by \cite{christensenEstimatingStabilityPsychological2021a} in the context of psychometric networks. 
A stability cut-off between $0$ and $1$ is then selected, such that nodes with a membership proportion below this threshold are considered insufficiently stable to be reliably assigned to a single community. These nodes can subsequently be treated as \textit{excluded nodes} in later stages of the analysis.

\section{Package overview} \label{sec:package}

\begin{figure}[t!]
\centering
\includegraphics{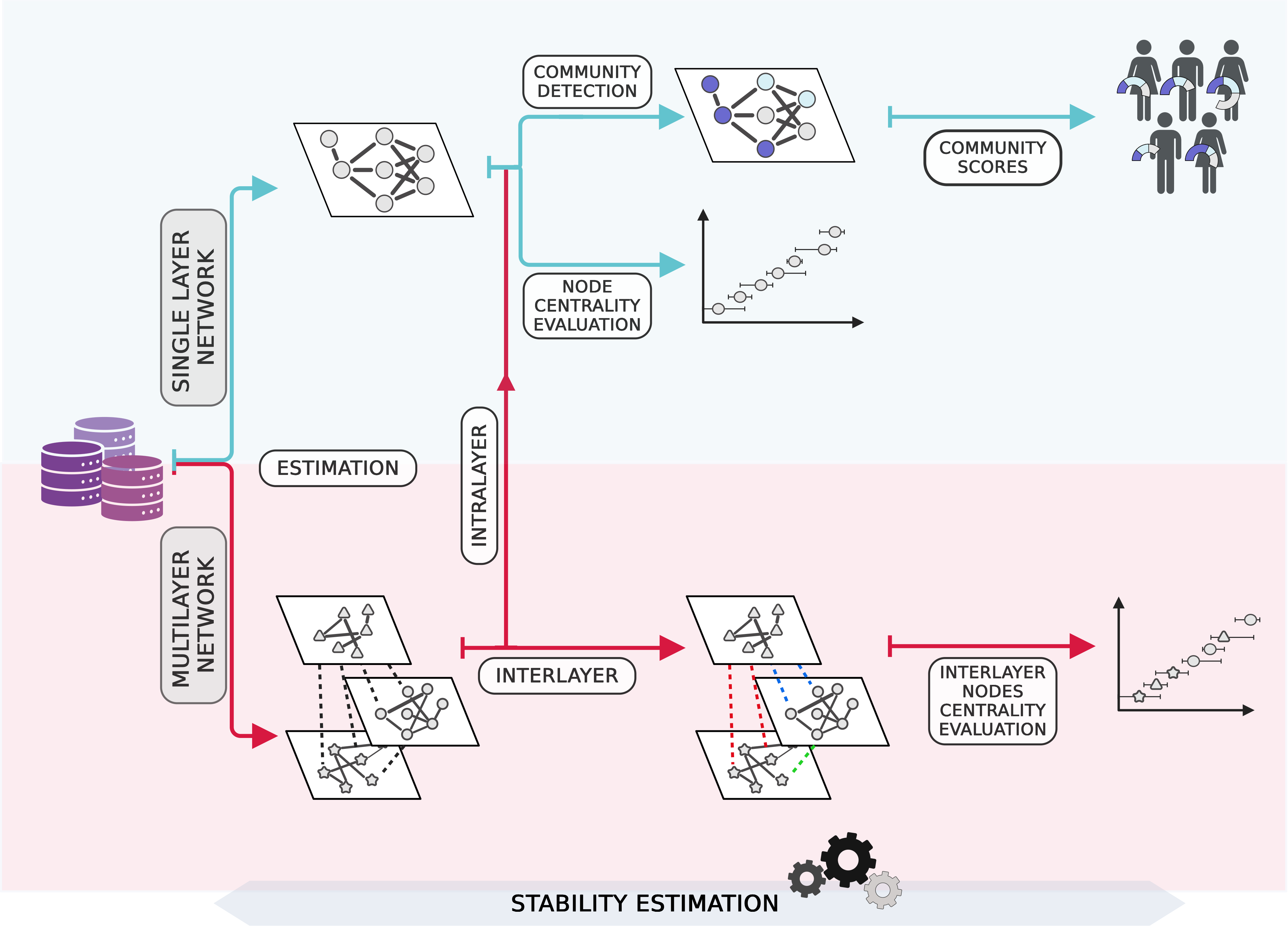}
\caption{\label{fig:WorkflowMixMashNet} Structure and functionalities of the \pkg{MixMashNet} package. }
\end{figure}

The \pkg{MixMashNet} package is organized around four principal components that represent the core modules of network analysis: network estimation, community detection, node-level analysis through centrality measures, and uncertainty evaluation. A key contribution of the package is the integration of these components into a unified analytical workflow, applicable to both single layer and multilayer cases (Figure~\ref{fig:WorkflowMixMashNet}).

The two core functions of the package are: \code{mixMN()}, for single layer networks, and \code{multimixMN()}, for multilayer networks. The two functions provide a wrapper for the four principal components of the network analysis, both relying on MGMs.
The multilayer extension preserves the same modeling framework while introducing additional arguments—\code{layers} and \code{layer_rules}—that allow users to specify the multilayer structure and define admissible connections between layers. 

The two functions return objects of class `\code{mixMN_fit}' and `\code{multimixMN_fit}', respectively. These objects collect all the main information about the estimated network, including the matched function call, the model settings, and information about the input data. In addition, they store the detected community structure and node-level centrality measures, as well as interlayer quantities in the multilayer case. Table~\ref{tab:methods} provides an overview of the available \proglang{S3} methods for objects of class `\code{mixMN_fit}' and `\code{multimixMN_fit}'.

\begin{table}[t!]
\centering
\begin{tabular}{lp{3cm}p{7cm}}
\hline
\textbf{Method}   & \textbf{Class}    & \textbf{Description} \\ \hline
\code{print()}   & `\code{mixMN_fit}', `\code{multimixMN_fit}' & Provides a compact overview of the fitted object, including model, graph, and estimation details. \\ 
\code{summary()} & `\code{mixMN_fit}', `\code{multimixMN_fit}' & Returns the top 10 intralayer/interlayer edges for fitted objects.  \\
\code{plot()}    & `\code{mixMN_fit}', `\code{multimixMN_fit}' &  Plots the network, node-level centrality indices or edge weights with bootstrap quantile regions, and node stability within communities. \\ 
\code{get_centrality()}  & `\code{mixMN_fit}', `\code{multimixMN_fit}'  & Extracts node-level centrality indices in a long-format data frame. \\ 
\code{get_edges()}       & `\code{mixMN_fit}', `\code{multimixMN_fit}' & Extracts edge-level summaries in a long-format data frame.  \\ 
\code{update_palette()}      & `\code{mixMN_fit}', `\code{multimixMN_fit}' & Updates the color palettes associated with communities and/or layers. \\ 
\code{layer_slice()}         &  `\code{multimixMN_fit}' &  Extracts a single layer from a multilayer object, returning a `\code{mixMN_fit}'.  \\ \hline
\end{tabular}
\caption{\label{tab:methods} Overview of methods associated with the `\code{mixMN\_fit}' and `\code{multimixMN\_fit}' objects.}
\end{table}

Alongside the core analytical functionalities, the \pkg{MixMashNet} package also offers an interactive \proglang{Shiny} application for the exploration of both single layer and multilayer networks.
The application can be directly launched from a fitted `\code{mixMN_fit}' or `\code{multimixMN_fit}' object, enabling dynamic inspection of the estimated network, community structure, and node-level indices through an interactive interface.
The Shiny application for single-layer network analysis is available at \url{https://arcbiostat.github.io/MixMashNet/articles/shiny-single.html}, while the corresponding application for multilayer networks is available at \url{https://arcbiostat.github.io/MixMashNet/articles/shiny-multilayer.html}.

\subsection{Network Estimation}

The estimation of the network is based on MGMs, which provide a flexible and generalized likelihood-based framework for handling data of different types. Especially, this framework does not rely on a single model, but rather encompasses a family of penalized regression models. Depending on the choice of tuning parameters, the underlying node-wise regressions can correspond to different regularization models, including lasso, ridge, or elastic net.
The \code{mixMN()} and \code{multimixMN()} functions retain all modeling options available in the \code{mgm} package. In particular, the parameter \code{lambdaSel} controls the selection of the penalization parameter and can be set to \code{"CV"} or \code{"EBIC"}. When cross-validation is used, \code{lambdaFolds} specifies the number of folds, whereas in the EBIC case \code{lambdaGam} indicates the $\gamma$ parameter.
Since penalization is formulated within an elastic net framework, the parameter \code{alphaSel} allows selecting $\alpha$ via \code{"CV"} or \code{"EBIC"}, while \code{alphaSeq} defines the sequence of candidate values in the interval $[0,1]$.

One of the strengths of network estimation in \pkg{MixMashNet} is the possibility to adjust for additional variables. These variables can be specified through the \code{covariates} argument as a character vector. They are included in the model estimation to account for potential confounding effects, but are not retained as nodes in the final network representation. Consequently, they are excluded from community detection and node-level centrality analyses.

\subsection{Community Detection}

Community detection is controlled through the \code{cluster_method} argument in both \code{mixMN()} and \code{multimixMN()}. This argument can be set to one of several built-in algorithms, including \code{"louvain"}, \code{"walktrap"}, \code{"infomap"}, and others, or provided as a user-defined function, allowing for flexible integration of alternative community detection strategies.
Additional arguments specific to the chosen method can be supplied via \code{cluster_args}.
\pkg{MixMashNet}, as previously described, also offers the possibility to exclude some nodes from the community detection step by specifying them in \code{exclude_from_cluster}, while the \code{treat_singletons_as_excluded} argument determines whether singleton communities are treated as excluded nodes in subsequent analyses.

From the detected communities, subject-level summaries can be obtained through the \code{community_scores()} function, which computes subject-level scores for each detected community. These scores can then be used in subsequent analyses.
The function requires a fitted object of class `\code{mixMN_fit}' or `\code{multimixMN_fit}', as well as the desired coverage level for the bootstrap quantile region of the scores (\code{quantile_level}). Optionally, a dataset can be specified via the \code{data} argument to enable external validation. External datasets must contain the same variables used for community assignment and be formatted consistently with the original model. If no dataset is provided, the original data used for model estimation is used by default, provided that \code{save_data = TRUE} was specified in \code{mixMN()} or \code{multimixMN()}. The computation of community scores relies on community loadings and is therefore available only when these have been estimated in the fitted model (i.e., when \code{compute_loadings = TRUE} in \code{mixMN()} or \code{multimixMN()}). When applied to single layer models, or to a specific layer of a multilayer model, the function returns an object of class `\code{community_scores}'. In the multilayer setting, if no layer is specified, a list of `\code{community_scores}' objects is returned, one for each layer where scores can be computed. The resulting objects are equipped with \code{print()} and \code{summary()} methods to facilitate interpretation.

\subsection{Node-level Centrality Measures}

In the \pkg{MixMashNet} framework, the main indices for evaluating node importance are available, including strength, closeness, betweenness, and expected influence. In the presence of communities, dedicated bridge centrality indices are available to assess the role of nodes as connectors between them. In the multilayer setting, corresponding indices are defined to capture cross-layer interactions.

Node-level centrality measures are computed within the \code{mixMN()} and \code{multimixMN()} functions and stored in the corresponding fitted objects. The \code{plot()} method provides a visual inspection of the selected indices, which can be specified through the \code{statistics} argument. Multiple indices can be displayed simultaneously (e.g., \code{statistics = c("strength", "expected_influence", "closeness", "betweenness")}).
For numerical inspection, the \code{get_centrality()} method extracts these measures in a long-format data frame, returning the original estimates. 
Furthermore, to support the interpretation of bridge centrality indices, the function \code{find_bridge_communities()} decomposes the bridge connectivity of a given node into contributions from different communities, identifying which communities drive its bridging role. The function requires a fitted object of class `\code{mixMN_fit}' or `\code{multimixMN_fit}', together with the \code{node} of interest and, in the multilayer setting, the corresponding \code{layer}. The function returns an object of class `\code{bridge_profiles}', with a dedicated \code{print()} method. In the multilayer context, the analogous function \code{find_bridge_layers()} extends this decomposition to interlayer analyses by quantifying the contribution of each layer to the interlayer connectivity of a node. This function requires a `\code{multimixMN_fit}' object, the \code{node} of interest, and the corresponding \code{layer}. It returns an object of class `\code{bridge_layer_profiles}', with a corresponding \code{print()} method.

\subsection{Uncertainty Evaluation}

One of the key features of \pkg{MixMashNet} is the ability to quantify the uncertainty of estimated network quantities through a non-parametric bootstrap procedure. The \code{boot_what} argument specifies which quantities should be resampled (by default, all supported quantities are included).
Uncertainty evaluation is integrated across all components of the analytical workflow. In particular, the stability of community assignments is assessed within the same bootstrap framework. Node stability can be visually inspected using the \code{plot()} method (with \code{what = "stability"}), while the \code{membershipStab()} function provides a more detailed, node-level quantification. The function returns an object of class `\code{membershipStab}', for which \code{print()}, \code{summary()}, and \code{plot()} methods are available.

Bootstrap results for centrality measures and edge weights can be explored within the same framework. In particular, the \code{plot()} method provides a unified visualization of bootstrap summaries, while methods such as \code{get_centrality()} and \code{get_edges()} allow numerical inspection of the corresponding quantities, including bootstrap means, standard errors, and quantile regions.

\section{Usage} \label{sec:usage}

The package \pkg{MixMashNet} is available on the Comprehensive \proglang{R} Archive Network (CRAN) at \url{https://CRAN.R-project.org/package=MixMashNet} and can be installed as follows:

\begin{Code}
R> install.packages("MixMashNet")
R> library("MixMashNet")
\end{Code}

In this section, we illustrate the usage of the \pkg{MixMashNet} package in its two main applications: single layer network and multilayer network estimation.

\subsection{Single Layer Network}

In this section, we describe how to estimate a single layer network using the \code{mixMN()} function, assess node membership stability, and compute centrality and bridge centrality indices, including metrics for excluded nodes.
Bootstrap quantile regions are computed for all these metrics, as well as for edge weights.
We also show how to obtain community scores with associated quantile regions.

We use the dataset \code{bacteremia}, included in the \pkg{MixMashNet} package. It consists of $16$ variables  measured in $7420$ patients with clinical suspicion of bacteremia who underwent blood culture testing at the Vienna General Hospital (2006–2010). The original dataset was introduced by \cite{ratzingerRiskPredictionModel2014} for the development of a screening model for bacteremia.
For network estimation, we considered $15$ variables: two demographic variables (\code{AGE} and \code{SEX}) and $13$ routinely collected laboratory biomarkers. 
These include white blood cell (\code{WBC}) and neutrophil counts (\code{NEU}), hemoglobin (\code{HGB}), platelet count (\code{PLT}), C-reactive protein (\code{CRP}), activated partial thromboplastin time (\code{APTT}), fibrinogen (\code{FIB}), creatinine (\code{CREA}), blood urea nitrogen (\code{BUN}), glucose (\code{GLU}), alanine aminotransferase (\code{ALAT}), total bilirubin (\code{GBIL}), and albumin (\code{ALB}).
The remaining variable, \code{BloodCulture}, is a binary indicator of bacteremia based on blood culture results. 
Although it is part of the original dataset, it is not included in the network estimation and is instead used in subsequent analyses, such as assessing associations with community scores derived from the estimated network.

\begin{Code}
R> data(bacteremia)
R> df <- bacteremia[, !names(bacteremia) 
\end{Code}

Input data must be provided as a data frame. Numeric variables are treated as
Gaussian; integer variables as Poisson unless they take only values in ${0,1}$,
in which case they are treated as binary categorical; and factor and logical
variables as categorical.

\subsubsection*{\textit{Network Estimation}}

We first illustrate how to estimate a single layer network.
As in \pkg{mgm}, prior to estimation we specify the method used to select the penalization parameter (\code{lambdaSel}), which in our setting is set to \code{"CV"}. We can also specify whether to scale the continuous variables (\code{scale = TRUE}). Additionally, \pkg{MixMashNet} introduces new parameters required for the functioning of the model. With \code{reps} we specify the number of bootstrap replications performed for assessing stability (set to \code{0} for no bootstrap estimation). In our example, we set \code{reps = 150}, which represents a commonly adopted compromise ensuring sufficient stability of edge-weight estimates while keeping the computational burden feasible \citep{epskampEstimatingPsychologicalNetworks2018}.
The argument \code{quantile_level} specifies the desired coverage level of the bootstrap quantile region (default \code{0.95}). The arguments \code{seed_model} and \code{seed_boot} allow fixing the random seed for the model estimation and the bootstrap replicates, respectively. In this example, the \code{"louvain"} algorithm is used for the \code{cluster_method} argument. The \code{covariates} argument is set to \code{c("AGE", "SEX")}. Finally, by setting \code{save_data = TRUE}, the original dataset used for model estimation is stored within the fitted object.

Since bootstrap-based procedures can be computationally intensive, \pkg{MixMashNet} is fully compatible with the parallel computing infrastructure provided by the \pkg{future} package \citep{bengtssonUnifyingFrameworkParallel2021}. A parallel backend can be specified (e.g., via \code{future::plan(multisession)}) to distribute bootstrap replications across multiple cores.

\begin{Code}
R> fit1 <- mixMN(data = df, scale = TRUE, lambdaSel = "CV", 
+       quantile_level = 0.95, reps = 150, seed_model = 42, seed_boot = 42, 
+       cluster_method = "louvain", covariates = c("AGE", "SEX"), 
+       save_data = TRUE)
\end{Code}

The output \code{fit1} is a `\code{mixMN_fit}' object. For an initial overview, we can simply call the \code{print()} method:

\begin{Code}
R> print(fit1)
MixMashNet fit
  Type: Single layer MGM (mixMN)
  Data: 7420 subjects x 15 variables
  Graph: 13 nodes, 50 edges
  Communities: 4 
  Covariates (adjusted for): AGE, SEX
  Lambda selection: CV
  Community detection: louvain
  Bootstrap replications: 150
  Bootstrapped quantities: general_index, bridge_index, excluded_index, 
                           community, loadings
  Data info:
    - Inferred as 'c' (categorical): SEX
\end{Code}

From this output we observe that the estimated network contains 13 nodes and 50 edges. Community detection using the Louvain algorithm identifies 4 communities. The printout also reports how many bootstrap replications were performed (here, 150) and which node- and community-level quantities were bootstrapped and stored in the \code{fit1} object. Finally, the "\code{Data info}" section summarizes the inferred MGM variable types, indicating that \code{SEX} was treated as a categorical variable in the model.

\begin{figure}[t!]
\centering
\includegraphics{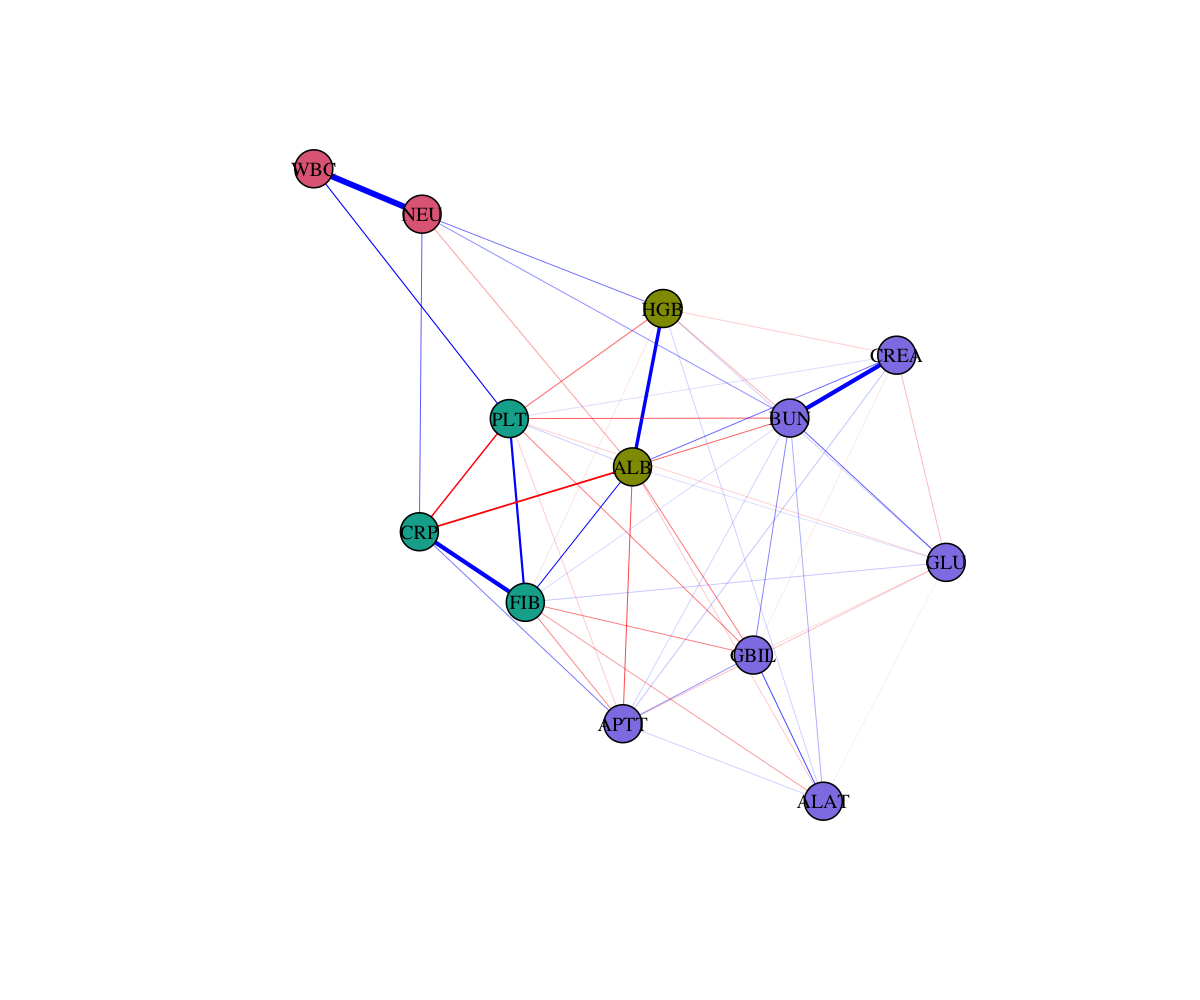}
\caption{\label{fig:fit1_plot} MGM network estimated on the \code{bacteremia} dataset. Nodes are arranged using the Fruchterman--Reingold layout. Blue edges indicate positive relationships, while red edges negative ones. The width of the edges is proportional to the absolute strength of the connection. Node colours represent the membership of the four identified communities.}
\end{figure}

We can also visualize the network by plotting the \code{fit1} object (Figure \ref{fig:fit1_plot}).

\begin{Code}
R> set.seed(28)
R> plot(fit1)
\end{Code}

 Nodes are arranged using the Fruchterman--Reingold layout \citep{fruchtermanGraphDrawingForcedirected1991}, a force-directed algorithm that positions nodes according to attractive and repulsive forces, such that strongly connected nodes are placed closer together. Nodes are colored according to the detected communities. Edges are shown in blue or red depending on whether the estimated association is positive or negative, and their widths are proportional to the absolute edge weights. The first community groups \code{WBC} and \code{NEU}; the second links \code{HGB} with \code{ALB}; the third includes \code{PLT}, \code{CRP} and \code{FIB}; and the fourth comprises  \code{APTT}, \code{CREA}, \code{BUN}, \code{GLU}, \code{ALAT} and \code{GBIL}.

\subsubsection*{\textit{Edge Weights}}

To further inspect the network, \pkg{MixMashNet} allows users to visualize and summarize edge weights together with their bootstrap quantile regions.
Edge-weight visualization can be obtained by calling the \code{plot()} method with \code{statistics = "edges"}, optionally specifying the number of edges to display via the \code{edges_top_n} argument, which ranks edges by the absolute value of their estimated weights.
Here, however, we focus on the numerical summary of edge weights.
This can be obtained using the \code{get_edges()} method.
In what follows, we report the top 10 edges ranked by the absolute value of their estimated weights:

\begin{Code}
R> get_edges(fit1) |> print(top_n = 10)

Edge-level summaries (intralayer):

  Layer: 1
       edge estimated mean         SE         95
                                              lower bound  upper bound
                      (bootstrap) (bootstrap) (bootstrap)  (bootstrap)
   WBC--NEU     0.951       0.952       0.019        0.908        0.975
  CREA--BUN     0.686       0.684       0.013        0.658        0.708
   CRP--FIB     0.681       0.683       0.007        0.669        0.698
   HGB--ALB     0.551       0.552       0.009        0.536        0.574
   PLT--FIB     0.354       0.357       0.013        0.330        0.384
   CRP--ALB    -0.286      -0.288       0.011       -0.311       -0.266
   PLT--CRP    -0.234      -0.238       0.011       -0.261       -0.218
   WBC--PLT     0.177       0.224       0.086        0.123        0.428
   FIB--ALB     0.162       0.163       0.013        0.137        0.187
 ALAT--GBIL     0.116       0.123       0.031        0.072        0.185   
... showing 10 of 50 rows after top_n filtering
\end{Code}

The output reports the estimated edge weights together with their bootstrap mean, standard error, and $95\%$ quantile regions.
Consistent with the network visualization, the strongest positive association is observed between \code{WBC} and \code{NEU}, characterized by a large estimated weight and a very narrow quantile region.
Strong positive associations are also found between \code{CREA} and \code{BUN}, as well as between \code{CRP} and \code{FIB}.
Negative associations are observed, for example, between \code{CRP} and \code{ALB}.

\subsubsection*{\textit{Node Membership Stability}}

In this section, we evaluate how consistently each node is assigned to the same community across bootstrap replications, and re-estimate the community structure including only stable nodes. To visually examine these stability values, we can use the \code{plot()} method:

\begin{Code}
R> plot(fit1, what = "stability")
\end{Code}

\begin{figure}[t!]
\centering
\includegraphics[width=\textwidth]{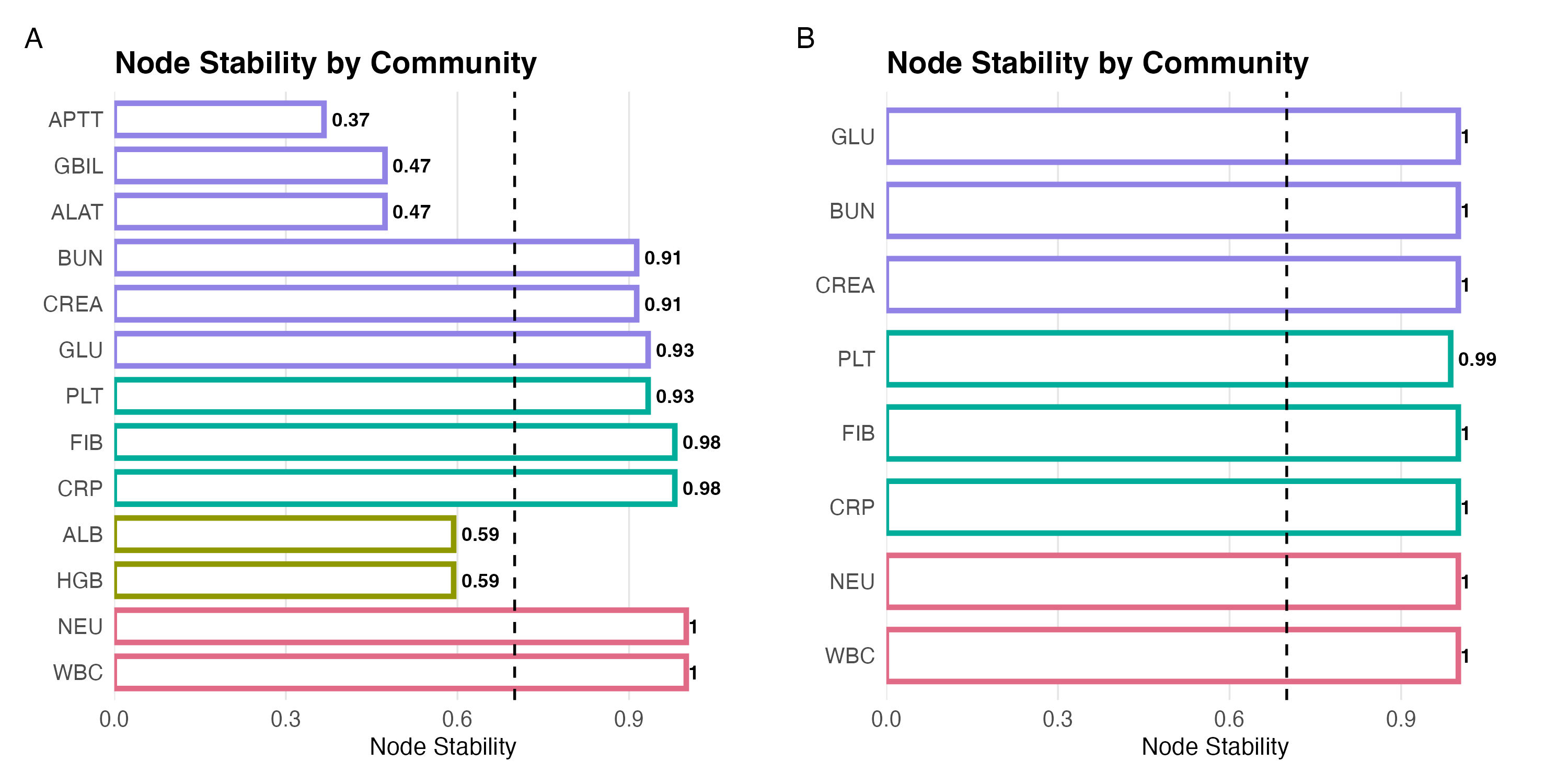}
\caption{\label{fig:fit_stability} Bootstrap-based node stability by community.
Panel~A shows node stability estimated from the initial network (\code{fit1}), while Panel~B reports node stability after excluding nodes with low stability (\code{fit2}).
Each bar corresponds to a node and is colored according to its community membership.
Stability values are displayed as labels at the end of each bar.
The dashed vertical line at 0.7 represents the reference threshold for acceptable stability.}
\end{figure}

In Figure~\ref{fig:fit_stability}A, each bar represents a node, colored according to its original community. Nodes are considered stable when the frequency of assignment to the same community across bootstrap replications is at least 0.7, following the recommendations of \cite{christensenEstimatingStabilityPsychological2021a}. The stability threshold can be modified via the \code{cutoff} argument in the \code{plot()} method.

In our example, the following variables do not meet the stability threshold: \code{APTT}, \code{GBIL}, \code{ALAT}, \code{ALB}, \code{HGB}.

Hence, the same \code{mixMN()} model is re-estimated, retaining the low-stability nodes in the network but excluding them from the community detection procedure using the argument \code{exclude_from_cluster}. This ensures that the clustering is based solely on nodes that exhibit sufficiently stable community membership. Furthermore, we set \code{treat_singletons_as_excluded = TRUE}  to exclude singleton nodes from community-based analyses.

\begin{Code}
R> fit2 <- mixMN(data = df, scale = TRUE, lambdaSel = "CV", 
+       reps = 150, seed_model = 42, seed_boot = 42, 
+       cluster_method = "louvain", covariates = c("AGE", "SEX"),
+       exclude_from_cluster = c("APTT", "GBIL", "ALAT", "ALB", "HGB"),
+       treat_singletons_as_excluded = TRUE, save_data = TRUE)
\end{Code}

The refined clustering identified three communities. The corresponding community assignments are stored in the fitted object and can be accessed as follows:

\begin{Code}
R> fit2$communities$groups
 WBC  NEU  PLT  CRP  FIB CREA  BUN  GLU 
   1    1    2    2    2    3    3    3 
Levels: 1 2 3
\end{Code}

For ease of interpretation, the color palette associated with the refined communities is updated to match the
corresponding communities identified in the initial model. The update is performed using the method
\code{update_palette()}.

\begin{Code}
R> fit2 <- update_palette(fit2, 
+          community_colors = c("2" = "#00AD9A", "3" = "#9183E6"))
\end{Code}

To evaluate whether this strategy improved the community structure, we again inspect node stability:

\begin{Code}
R> plot(fit2, what = "stability")
\end{Code}

As shown in Figure ~\ref{fig:fit_stability}B, the stability of the remaining nodes improved markedly after excluding those with lower stability, leading to more coherent and reliable network communities.

\begin{figure}[t!]
\centering
\includegraphics{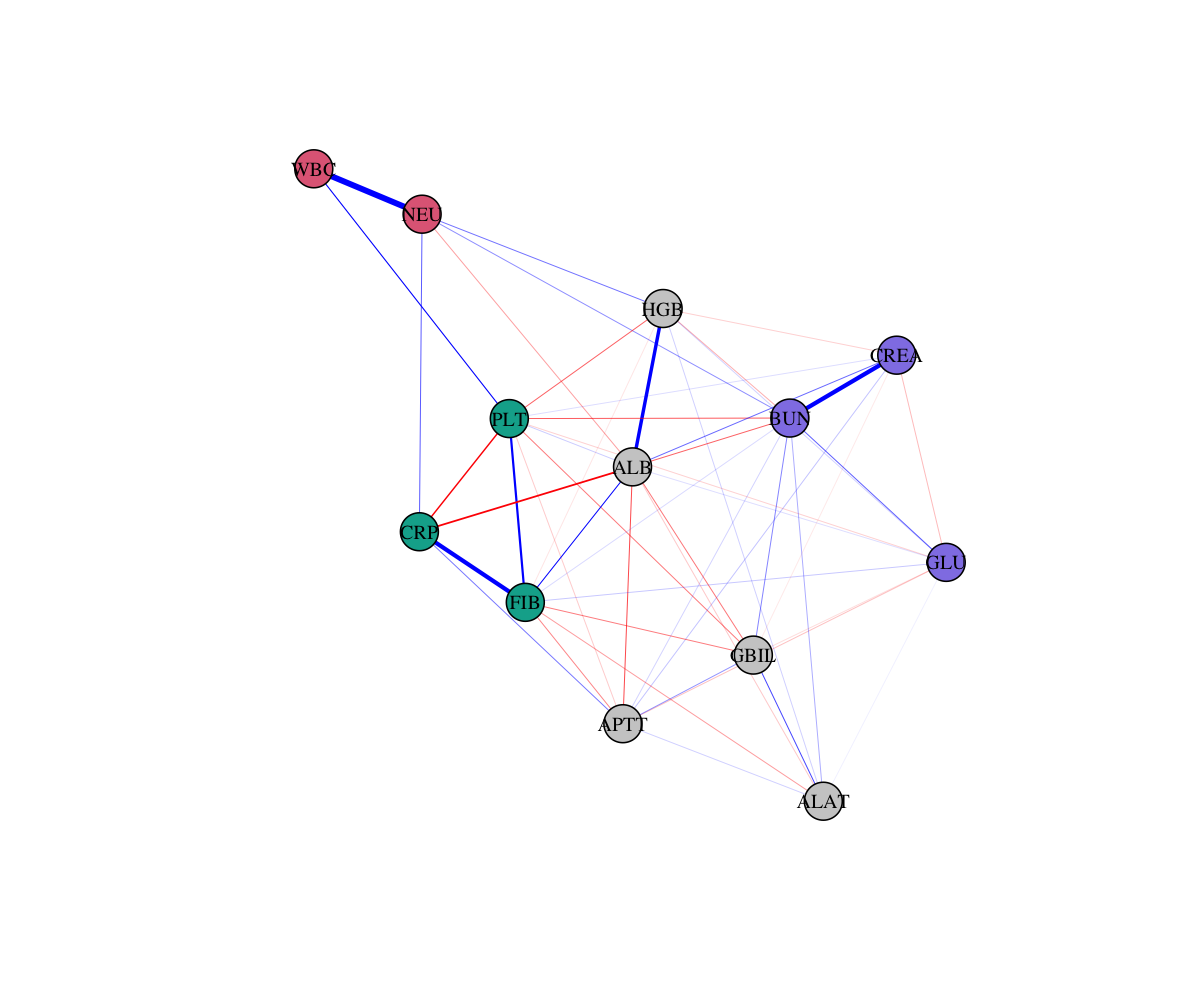}
\caption{\label{fig:fit2_plot} MGM network re-estimated on the \code{bacteremia} dataset after removing low-stability nodes. Nodes are arranged using the Fruchterman--Reingold layout. Blue edges indicate positive relationships, while red edges indicate negative ones. The width of each edge is proportional to the absolute strength of the corresponding connection. Node colours represent the membership of the three identified communities. Grey nodes are not assigned to any community.}
\end{figure}

We then visualize the re-estimated network with updated community assignments (Figure ~\ref{fig:fit2_plot}):

\begin{Code}
R> set.seed(28)
R> plot(fit2)
\end{Code}

In this case, nodes displayed in grey (i.e., excluded nodes) remain part of the network but are not considered in community-based interpretations.

\subsubsection*{\textit{Centrality and Bridging Analyses}}

Once node membership stability has been assessed, we can reliably interpret the structural role of each variable in the network, both within and across communities. In \pkg{MixMashNet}, this is achieved through a set of centrality and bridging indices.

\paragraph{General Centrality Indices.}

We can explore node importance using the \code{plot()} method by specifying the indices of interest via \code{statistics = c("strength", "expected_influence", "closeness", "betweenness")}.

\begin{Code}
R> plot(fit2, statistics =  c("strength", "expected_influence", 
+	"closeness", "betweenness"))
\end{Code}

For numerical values, the \code{get_centrality()} method can be used, again specifying the desired indices via the \code{statistics} argument.

\begin{figure}[t!]
\centering
\includegraphics{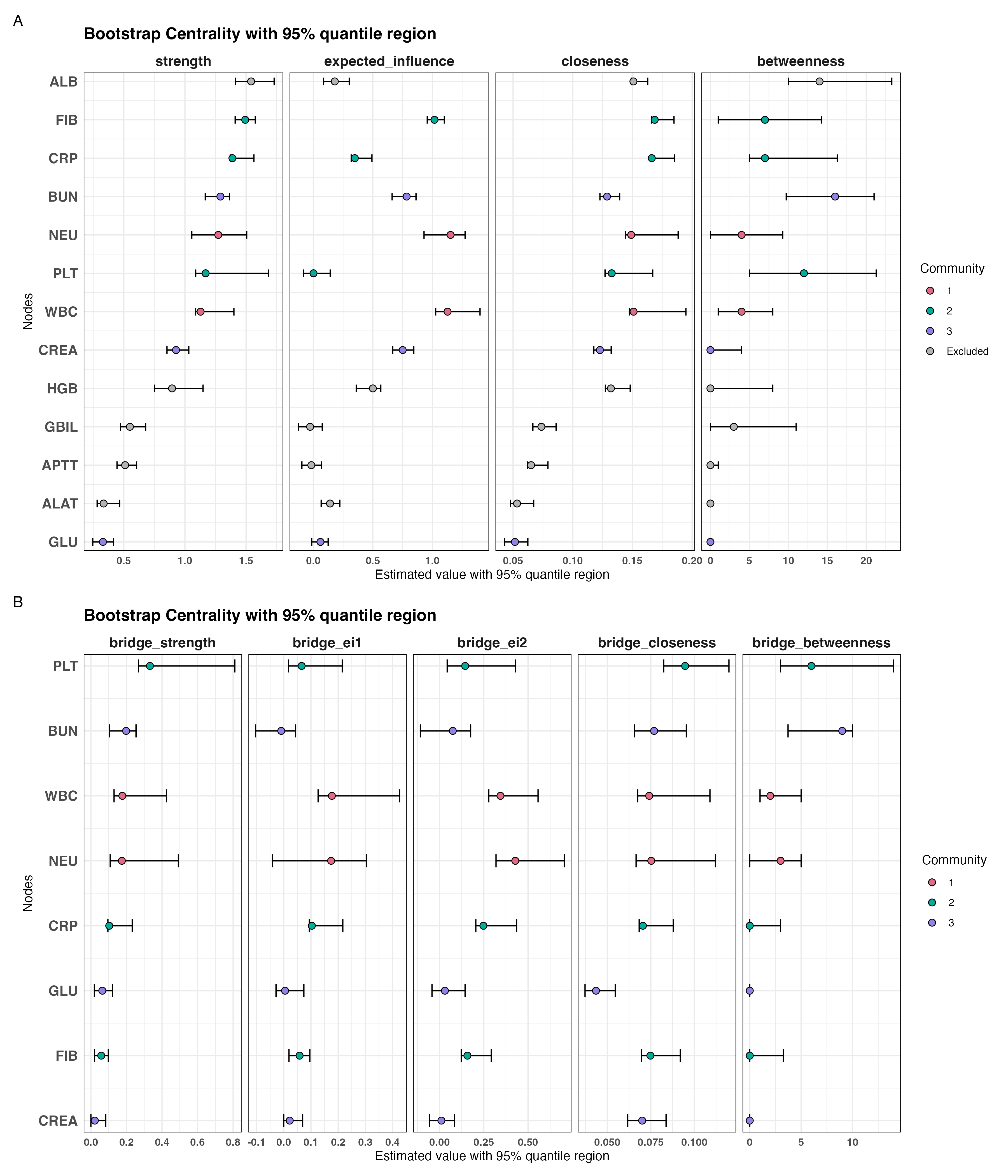}
\caption{\label{fig:fit2_centrality} Centrality estimates with 95\% quantile regions for the \code{fit2} object.
Panel~A reports general node-level centrality indices (strength, expected influence, closeness, and betweenness), while Panel~B shows bridge centrality indices with respect to the detected communities.
Nodes are ordered according to the strength index in Panel~A and according to the bridge strength index in Panel~B.}
\end{figure}

As shown in Figure~\ref{fig:fit2_centrality}A, nodes are ordered according to the strength index.
\code{ALB} exhibits the highest strength and comparatively high values of closeness and betweenness, indicating a central role both in terms of direct connectivity and overall proximity within the network. By contrast, for example, \code{APTT} and \code{ALAT} display consistently low centrality values, suggesting a more peripheral position in the network.

\paragraph{Bridge Centrality Indices}

An important next step is to evaluate how nodes act as connectors between communities. To this end, we focus on bridge-specific centrality indices — namely bridge strength, bridge expected influence (1- and 2-step), bridge closeness, and bridge betweenness. Only nodes assigned to a community are displayed in this type of plot.

\begin{Code}
R> plot(fit2, statistics = c("bridge_strength", "bridge_ei1", "bridge_ei2", 
+ "bridge_closeness", "bridge_betweenness"))
\end{Code}

As shown in Figure~\ref{fig:fit2_centrality}B, \code{PLT} has the highest bridge strength, followed by \code{BUN}, suggesting that these markers may serve as key connectors between communities. Conversely, nodes such as \code{GLU} and \code{CREA} show low bridge centrality, indicating weaker cross-community influence.

When investigating bridging nodes, a natural question is which communities are most strongly connected to a given node, excluding its own. To address this, we use the \code{find_bridge_communities()} function. In particular:

\begin{Code}
R> bridge <- find_bridge_communities(fit2, "PLT")
R> print(bridge, statistic = "bridge_strength")

Bridge Strength
============================================================ 

Overall:
   0.332 

By community:
# A tibble: 2 × 2
  Community `Contribution (|w|)`
      <dbl>                <dbl>
1         1                0.177
2         3                0.155
\end{Code}

This output suggests that \code{PLT}’s bridging role is mainly driven by its connections with nodes in community 1, while community 3 contributes to a lesser extent. This information enhances the interpretation of bridge centrality metrics by revealing which group-to-group interactions a node is responsible for.

\paragraph{Bridge Centrality Indices for excluded nodes.}
Nodes that were excluded from clustering can still play a role in facilitating communication across communities. For this reason, \pkg{MixMashNet} provides bridge centrality metrics computed specifically for these nodes. These include \code{"bridge_strength_excluded"}, \code{"bridge_ei1_excluded"}, \code{"bridge_ei2_excluded"}, \code{"bridge_closeness_excluded"}, and \code{"bridge_betweenness_excluded"}. They can be specified within the \code{statistics} argument of \code{plot()}.

In this case, we show the numerical summary of the bridge closeness index for excluded nodes using the \code{get_centrality()} method.

\begin{Code}
R> get_centrality(fit2, statistics = "bridge_closeness_excluded")

Node-level centrality indices (intralayer):

  Metric: bridge_closeness_excluded 
 node estimated mean        SE          95
                                        lower bound  upper bound
                (bootstrap) (bootstrap) (bootstrap)  (bootstrap)
 ALAT     0.047       0.050       0.005        0.042        0.061
  ALB     0.096       0.103       0.007        0.090        0.116
 APTT     0.054       0.060       0.004        0.054        0.069
 GBIL     0.064       0.067       0.006        0.055        0.078
  HGB     0.087       0.089       0.007        0.078        0.105
\end{Code}

The excluded variables differ in the extent to which they may influence communication between communities. In particular, \code{ALB} displays the highest closeness values, with a narrow quantile region, suggesting a potentially relevant cross-community role. Conversely, markers such as \code{APTT} and \code{ALAT} show low bridging influence.
Overall, this analysis shows that excluding nodes from clustering does not necessarily imply marginal importance: some may still function as potential connectors in the system. To further clarify these roles, one can again use the \code{find_bridge_communities()} function to determine which community links mainly contribute to each excluded node’s bridging capability.

\subsubsection*{\textit{Community Scores}}

This section illustrates how to compute community scores from the fitted network. To this end, we use the \code{community_scores()} function with \code{quantile_level = 0.95}:

\begin{Code}
cs <- community_scores(fit = fit2, quantile_level = 0.95)
\end{Code}

\begin{figure}[t!]
\centering
\includegraphics[width=1\textwidth]{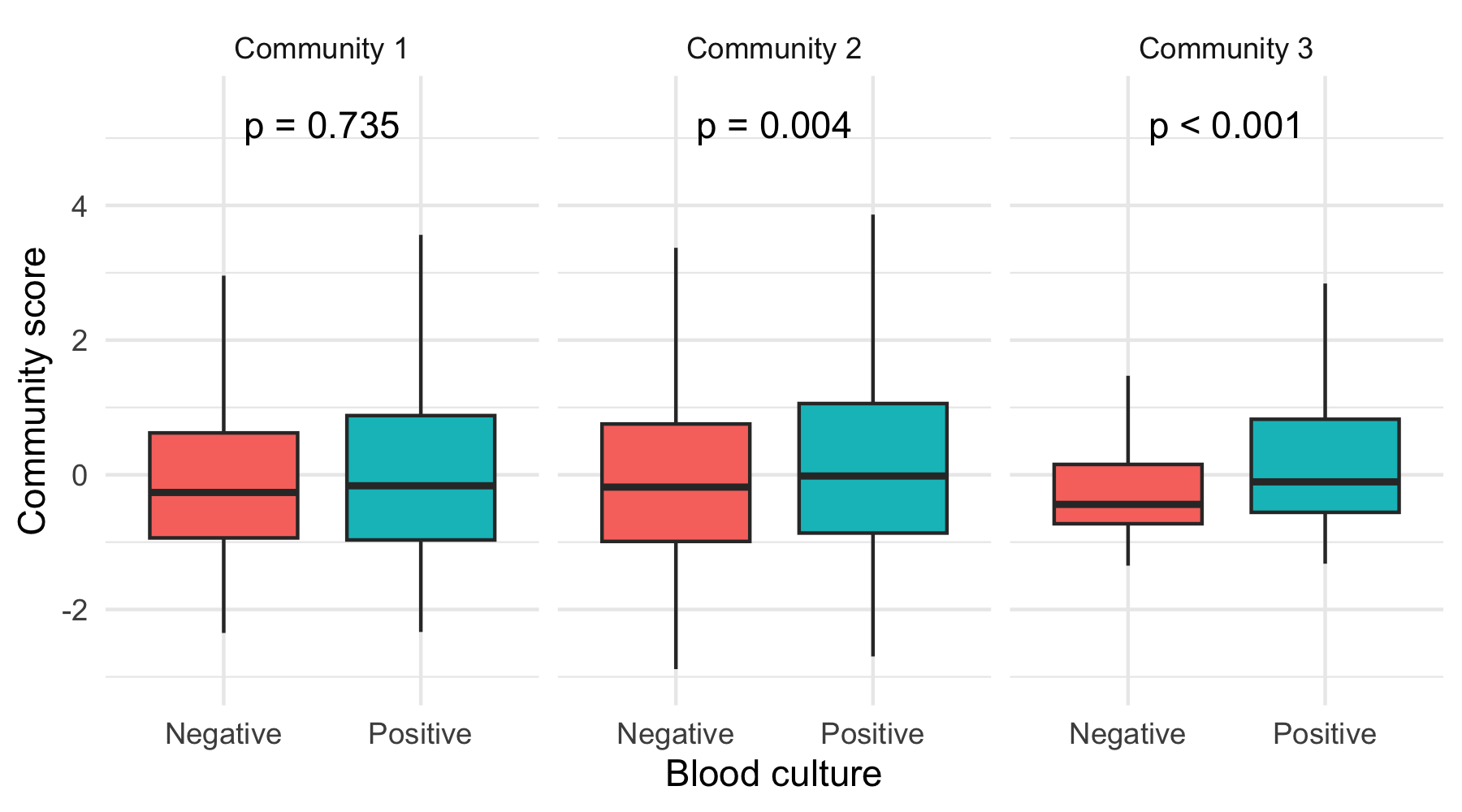}
\caption{\label{fig:box_plot} Box plots summarize the distribution of the community scores for blood culture–negative and blood culture–positive participants in Communities 1, 2 and 3. Between-group differences were assessed using the Wilcoxon rank-sum test, with p-values adjusted for multiple comparisons using the Bonferroni correction.
}
\end{figure}

The resulting scores are stored in \code{cs$scores}. Each score represents a standardized, subject-level summary of the variables within a given community, taking into account their interdependence.

As an illustrative example, we focus on comparing the three scores between patients with negative and positive blood culture results. The distributions across the two groups are summarized using boxplots. Figure~\ref{fig:box_plot} shows that the community~1 score (\code{WBC} and \code{NEU}) does not differ between the two groups ($p=0.735$), while for community~2 (\code{CRP}, \code{FIB}, and \code{PLT}) and community~3 (\code{CREA}, \code{BUN}, and \code{GLU}), participants with a positive blood culture result had, on average, higher scores compared with those with a negative result ($p=0.004$ and $p<0.001$, respectively). This suggests that these communities may play a role in differentiating patients based on their infection status.

\subsection{Multilayer Network}

In this section, we illustrate how to estimate a multilayer network using the \code{multimixMN()} function, focusing on the evaluation of interlayer edges and node-level centrality indices.
The assessment of intralayer edges, intralayer centrality indices, membership stability and community scores follows the same workflow as in the single layer case, and is therefore not repeated here.

We rely on the \code{nhanes} dataset included in the \pkg{MixMashNet} package, which contains $2759$ subjects and $29$ variables derived from the National Health and Nutrition Examination Survey (NHANES) \citep{centersfordiseasecontrolandpreventioncdc&nationalcenterforhealthstatisticsnchsNationalHealthNutrition2020}. A multilayer network with three layers is specified. The first layer represents biochemical biomarkers and includes twelve continuous variables: total cholesterol (\code{TotChol}), high-density lipoprotein cholesterol (\code{HDL}), creatinine (\code{Creatinine}), uric acid (\code{UricAcid}), alanine aminotransferase (\code{ALT}), aspartate aminotransferase (\code{AST}), gamma-glutamyl transferase (\code{GGT}), bilirubin (\code{Bilirubin}), albumin (\code{Albumin}), total protein (\code{TotProtein}), glycated hemoglobin (\code{HbA1c}), and high-sensitivity C-reactive protein (\code{hsCRP}). The second layer contains seven continuous anthropometric measures: body mass index (\code{BMI}), waist circumference (\code{Waist}), height (\code{Height}), arm circumference (\code{ArmCirc}), hip circumference (\code{HipCirc}), leg length (\code{LegLength}), and arm length (\code{ArmLength}). The third layer comprises seven lifestyle and behavioral variables: trouble sleeping (\code{TroubleSleep}), physical activity (\code{PhysicalActivity}), smoking (\code{Smoke}), recreational drug use (\code{Drug}), dietary quality (\code{Diet}), alcohol consumption (\code{Alcohol}), and working status (\code{Work}).

The multilayer network is adjusted for household monthly income (\code{MonInc}), categorized into twelve ordered groups according to NHANES coding, as well as for age (\code{Age}) and sex (\code{Gender}).

\begin{Code}
R> data(nhanes)

R> bio_vars  <- c("TotChol", "HDL", "Creatinine", "UricAcid", "ALT", "AST",
+		"GGT", "Bilirubin", "Albumin", "TotProtein", 
+		"HbA1c", "hsCRP")
R> ant_vars  <- c("BMI", "Waist", "Height", "ArmCirc", "HipCirc",
+		"LegLength", "ArmLength") 
R> life_vars <- c("TroubleSleep", "PhysicalActivity", "Smoke", "Drug",
+		"Diet", "Alcohol", "Work")
\end{Code}

\subsubsection*{\textit{Network Estimation}}

This section describes the estimation of the multilayer network. 
The arguments required to estimate the model are the same as for the \code{mixMN()} function. Here, we estimate the model using cross-validation to select the regularization parameter (\code{lambdaSel = "CV"}), with \code{reps = 150} bootstrap replications and the Walktrap algorithm for community detection. The model is adjusted for \code{"Age"}, \code{"Gender"}, and \code{"MonInc"}, which are excluded from the network and treated as covariates.
Additionally, \code{multimixMN()} requires a \code{layers} vector specifying the layer membership of each variable included in the network and argument \code{layer_rules}, which is an $L \times L$ matrix (with $L$ the number of layers) whose entries are either 1 or 0: a value of 1 indicates that connections between the corresponding pair of layers are allowed, whereas 0 denotes that no connections are permitted. Unspecified off-diagonal entries are treated as 0, diagonal entries are set to 1 by default, and the matrix is assumed to be symmetric, so it is sufficient to define one direction for each allowed connection. In this example, connections are allowed between all pairs of layers.

\begin{Code}
R> layers <- setNames(c(rep("bio", length(bio_vars)),
+                       rep("ant", length(ant_vars)),
+                       rep("life", length(life_vars))),
+                     c(bio_vars, ant_vars, life_vars))
R> layer_names <- c("bio", "ant", "life")
R> layer_rules <- matrix(NA, nrow = 3, ncol = 3,
+                        dimnames = list(layer_names, layer_names))
R> layer_rules["bio",  "ant"] <- 1
R> layer_rules["ant", "life"] <- 1
R> layer_rules["life", "bio"] <- 1
\end{Code}

We can then use the \code{multimixMN()} function to estimate the multilayer network:

\begin{Code}
R> nhanesMulti1 <- multimixMN(data = nhanes, scale = TRUE,
+               layers = layers, layer_rules = layer_rules,
+               quantile_level = 0.95, reps = 150, lambdaSel = "CV",  
+               seed_model = 42, seed_boot = 42, 
+               cluster_method = "walktrap", 
+               covariates = c("Age", "Gender", "MonInc"))
\end{Code}

The function \code{multimixMN()} returns an object of class `\code{multimixMN_fit}'. A general overview of the fitted multilayer network can be obtained using the \code{print()} method:

\begin{Code}
R> print(nhanesMulti1)
MixMashNet fit
  Type: Multilayer MGM (multimixMN)
  Data: 2759 subjects x 29 variables
  Layers (3):
    - bio: 12 nodes, 40 edges
    - ant: 7 nodes, 16 edges
    - life: 7 nodes, 17 edges
  Interlayer edges:
    - bio_ant: 8 edges
    - bio_life: 30 edges
    - ant_life: 10 edges
  Graph: 26 nodes, 121 edges
  Communities per layer:
    - bio: 2
    - ant: 2
    - life: 3
  Covariates (adjusted for): Age, Gender, MonInc
  Lambda selection: CV
  Community detection: walktrap
  Bootstrap replications: 150
  Bootstrapped quantities: general_index, interlayer_index, bridge_index, 
                           excluded_index, community, loadings
  Data info:
    - Inferred as 'c' (categorical): TroubleSleep, PhysicalActivity, Smoke, 
    Drug, Diet, Alcohol, Work, MonInc, Gender
\end{Code}

The estimated multilayer network includes 26 nodes and 121 edges, structured into three layers: 12 biological variables (40 intralayer edges), 7 anthropometric variables (16 edges), and 7 lifestyle variables (17 edges). Cross-layer connections result in a total of 48 interlayer edges, with 30 edges connecting the biological and lifestyle layers, 8 linking biological and anthropometric variables, and 10 linking anthropometric and lifestyle variables.
Community detection with the Walktrap algorithm identified two communities within both the biological and anthropometric layers, and three communities within the lifestyle layer.

We can then visualize the estimated multilayer network:

\begin{Code}
R> set.seed(123)
R> plot(nhanesMulti1)
\end{Code}

\begin{figure}[t!]
\centering
\includegraphics{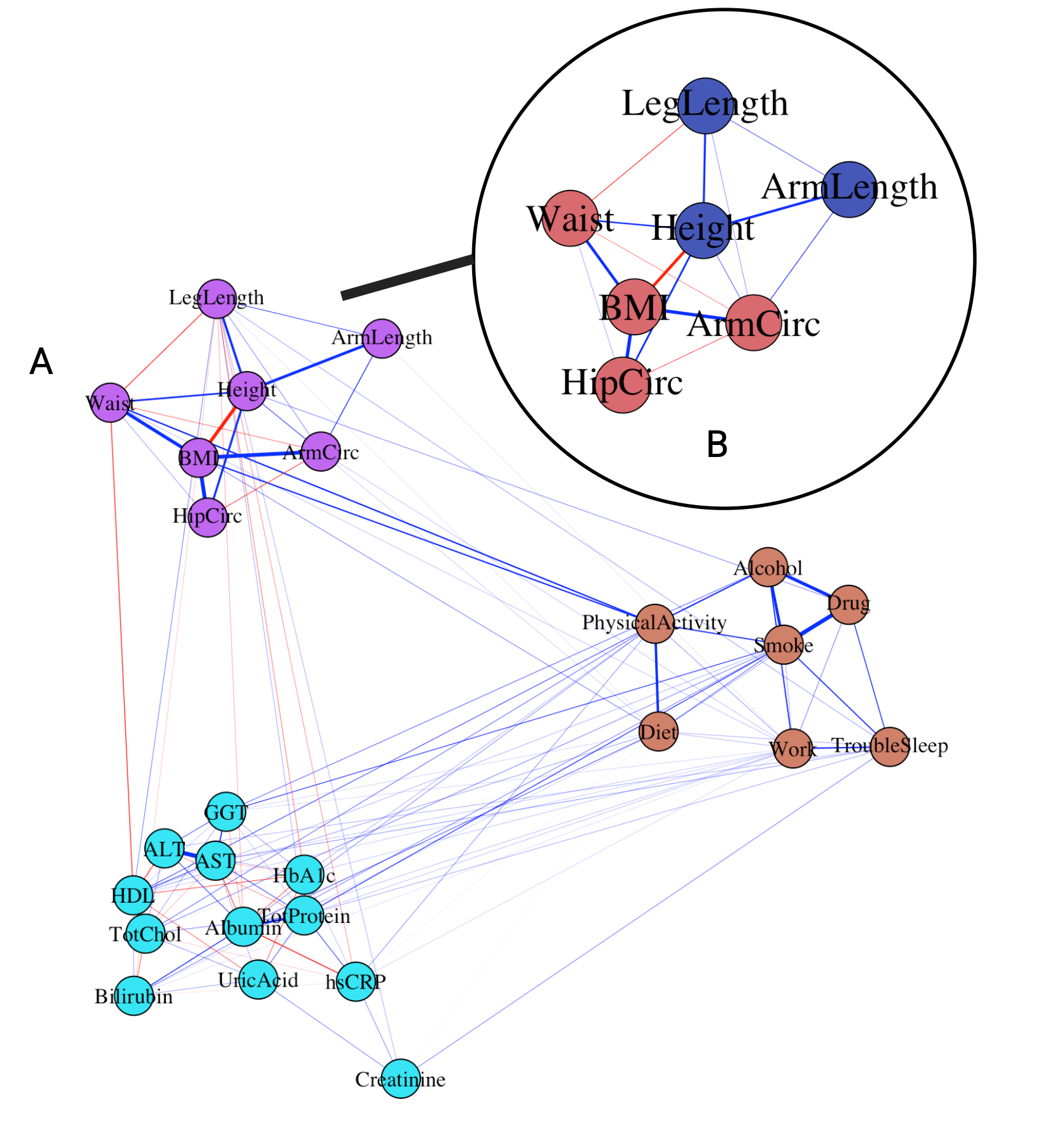}
\caption{\label{fig:multi_zoom} Multilayer MGM network estimated on the \code{nhanes} dataset. In panel A, the full multilayer structure is shown, with nodes arranged using the Fruchterman–Reingold layout. Edge width is proportional to the absolute strength of the estimated conditional associations, with blue and red edges indicating positive and negative relationships, respectively. Node colours identify the three layers: biological (turquoise), anthropometric (purple), and lifestyle (orange). Panel B presents a magnified view of the anthropometric layer, highlighting communities. In this panel, node colours represent community membership within the anthropometric layer, while edge colour and width retain the same interpretation as in panel A.}
\end{figure}

Figure \ref{fig:multi_zoom}A shows the final multilayer graph in a two-dimensional layout. Nodes are positioned to maximize visual separation between layers while preserving network topology. The colour-coding highlights the three layers included in the model, providing an intuitive depiction of both within and cross-layer associations. Turquoise nodes belong to the biological layer, purple nodes to the anthropometric layer, and orange nodes to the lifestyle layer. Cross-layer connections are more prominent between the biological and lifestyle layers, whereas links involving the anthropometric layer appear less frequent.

For the visualization of community assignments within each layer, we can again use the \code{plot()} method and specify the argument \code{color_by = "community"}. However, as an illustration of how individual layers can be accessed and visualized from a multilayer object, we focus here on the anthropometric layer:

\begin{Code}
R> set.seed(12)
R> plot(nhanesMulti1, layer = "ant")
\end{Code}

Figure~\ref{fig:multi_zoom}B illustrates the community structure of the anthropometric layer. One community groups together the length-related measures (\code{LegLength}, \code{Height}, \code{ArmLength}), whereas the second community includes circumferential and adiposity measures (\code{BMI}, \code{Waist}, \code{ArmCirc}, \code{HipCirc}).

\subsubsection*{\textit{Interlayer Edges}}

One of the main interests in multilayer networks lies in examining the connections between different layers. In \pkg{MixMashNet}, these interlayer associations can be visualized using the \code{plot()} function.
In the \code{nhanes} example, we display the ten strongest interlayer edges, ranked by absolute weight across all interlayer connections and subsequently grouped by layer. To focus on interlayer quantities, we specify \code{what = "inter"} and request edge statistics via \code{statistics = "edges"}:

\begin{Code}
R> plot(nhanesMulti1, what = "inter", statistics = "edges", edges_top_n = 10) 
\end{Code}

\begin{figure}[t!]
\centering
\includegraphics{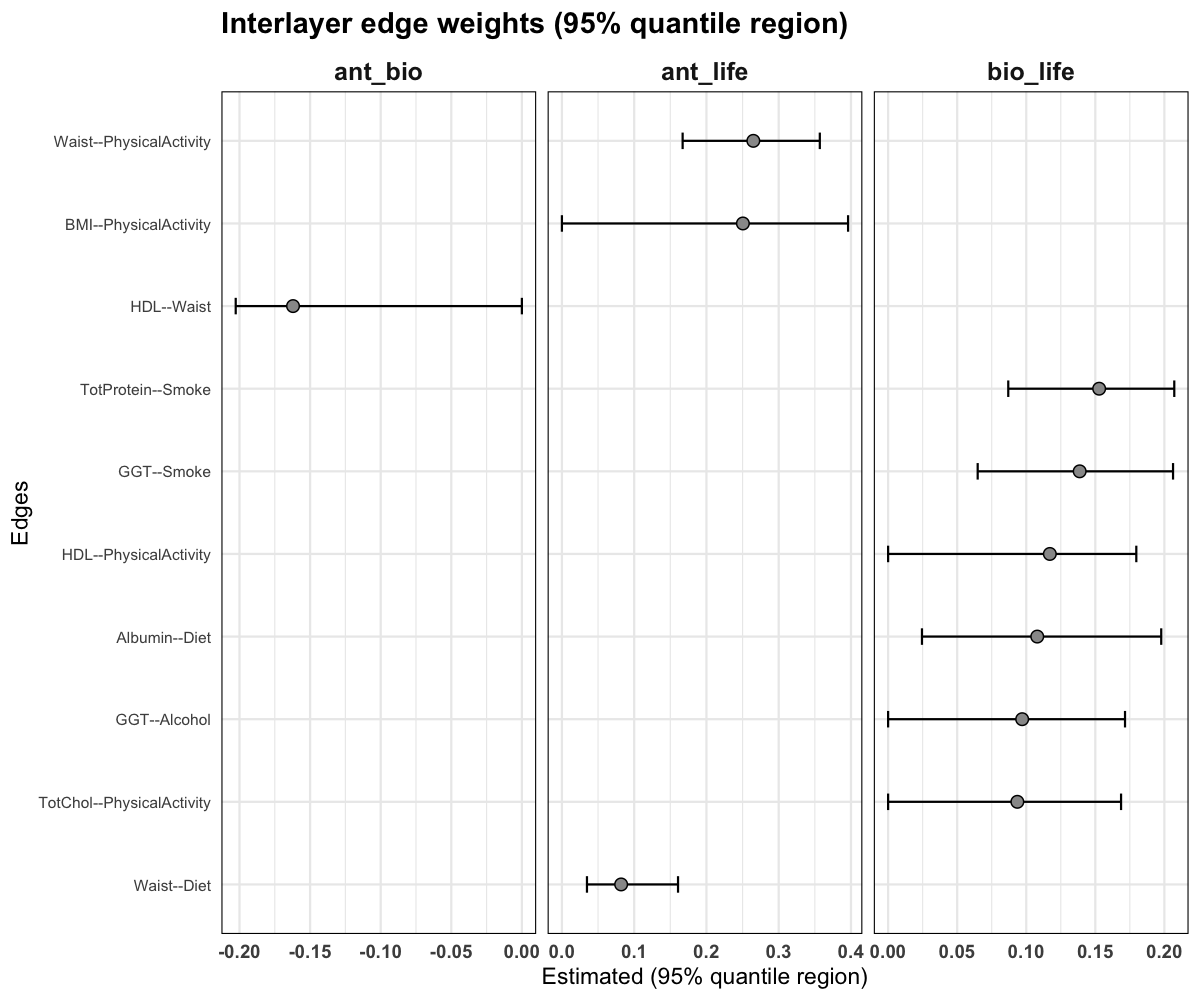}
\caption{\label{fig:edges_inter_plot} Estimated interlayer edge weights with 95\% quantile regions for the \code{nhanesMulti1} object. The figure displays the ten strongest interlayer
edges overall, ranked by absolute weight across all interlayer connections and subsequently grouped by layer. The left panel displays interlayer edges between the biological and anthropometric layers, central panel between the lifestyle and anthropometric layers and right panel between biological and lifestyle layers.}
\end{figure}

Figure~\ref{fig:edges_inter_plot} shows the top-10 estimated interlayer edges together with their 95\% quantile regions. The left panel shows connections between biological and anthropometric variables, the central panel illustrates links between anthropometric and lifestyle variables, and the right panel displays connections between the biological and lifestyle layers.
Overall, biological variables (e.g., \code{TotProtein}, \code{GGT}, \code{HDL}) show positive associations with lifestyle measures such as \code{Smoke} and \code{PhysicalActivity}. Lifestyle-related variables (e.g., \code{PhysicalActivity}, \code{Diet}) exhibit connections with anthropometric measures such as \code{Waist} and \code{BMI}. 

\subsubsection*{\textit{Interlayer Indices}}

\pkg{MixMashNet} also provides tools to visualize interlayer centrality indices. Such visualizations can be obtained by specifying the desired metric in the \code{statistics} argument of the \code{plot()} function, choosing one of \code{"strength"}, \code{"expected_influence"}, \code{"closeness"}, or \code{"betweenness"}.

The indices can also be inspected numerically using the \code{get_centrality()} function, which returns interlayer centrality estimates together with their bootstrap-based uncertainty for multilayer objects.
Specifically, to focus on the top 10 nodes ranked by interlayer betweenness, we specify \code{what = "inter"} and \code{statistics = "betweenness"}:

\begin{Code}
R> get_centrality(nhanesMulti1, what = "inter", statistics = "betweenness") |> 
+  print(top_n = 10)

Node-level centrality indices (interlayer-only graph):

  Metric: betweenness 
      node layer estimated mean        SE          95
                           (bootstrap) (bootstrap) lower bound  upper bound 
                                                   (bootstrap)  (bootstrap)
 PhysicalA  life        57      63.453      30.013       14.000      138.200
     Smoke  life        54      44.927      24.599        9.725      104.100
 LegLength   ant        52      35.520      21.257        4.000       85.550
  TroubleS  life        48      35.200      23.598        0.000       86.200
      Diet  life        43      37.967      23.286        1.000       89.650
      Work  life        43      49.453      26.936        2.725       99.825
   TotProt   bio        31      25.980      18.810        0.000       67.275
     Waist   ant        24      33.773      23.665        0.000       85.550
   Albumin   bio        24      12.487      14.797        0.000       49.375
    Height   ant        22      10.367      13.575        0.000       46.000                      
... showing 10 of 26 rows after top_n filtering

\end{Code}

As shown above, \code{PhysicalActivity}, followed by \code{Smoke}, exhibits the highest interlayer betweenness, indicating a prominent bridging role between layers. This finding highlights the central importance of lifestyle-related variables in mediating connections between biological functioning and anthropometric measures across layers.

\section{Summary} \label{sec:summary}

In this paper, we have presented the \proglang{R} package \pkg{MixMashNet}, which provides a unified framework for estimating and analyzing single and multilayer networks using MGMs. The package accommodates any data types and, through regularization-based MGM estimation, is suitable for high-dimensional settings. It enables users to impose a predefined multilayer topology, and offers bootstrap-based quantile regions for both edge weights and node-level indices. In addition, it evaluates the stability of node membership within communities and derives community scores that summarize the latent dimensions revealed by network clustering. Furthermore, \pkg{MixMashNet} extends traditional node-centrality analysis by incorporating measures that quantify how nodes mediate communication across communities and layers, including dedicated metrics for nodes that are not assigned to any community. Two \proglang{Shiny} applications are also available, providing an interactive exploration of the networks estimated.

Current limitations of \pkg{MixMashNet} include its exclusive reliance on MGMs, which restricts the ability to model non-linear relationships. The package is actively maintained, and future developments will include support for alternative network estimators to broaden the range of structural assumptions that can be explored.

\bibliography{refs}

\newpage

\end{document}